# Structure-property relationships of elastin-like polypeptides – a review of experimental and computational studies

Diego López Barreiro[1*], Inge J. Minten[2], Jens C. Thies[3], Cees M. J. Sagt[1]

[1]*DSM Biotechnology Center, DSM, Alexander Fleminglaan 1, 2613 AX Delft, The Netherlands*

[2]*DSM Materials Science Center - Applied Science Center, DSM, Urmonderbaan 22, 6160 BB, Geleen, The Netherlands*

[3]*DSM Biomedical, DSM, Koestraat 1, 6167 RA, Geleen, The Netherlands*

*corresponding author, Diego.Lopez-Barreiro@dsm.com*

**Abstract**

Elastin is a structural protein with outstanding mechanical properties (e.g., elasticity and resilience) and biologically relevant functions (e.g., triggering responses like cell adhesion or chemotaxis). It is formed from its precursor tropoelastin, a 60-72 kDa water-soluble and temperature-responsive protein that coacervates at physiological temperature, undergoing a phenomenon termed lower critical solution temperature (LCST). Inspired by this behaviour, many scientists and engineers are developing recombinantly produced elastin-inspired biopolymers, usually termed elastin-like polypeptides (ELPs). These ELPs are generally comprised of repetitive motifs with the sequence VPGXG, which corresponds to repeats of a small part of the tropoelastin sequence, X being any amino acid except proline. ELPs display LCST and mechanical properties similar to tropoelastin, which renders them promising candidates for the development of elastic and stimuli-responsive protein-based materials. Unveiling the structure-property relationships of ELPs can aid in the development of these materials by establishing the connections between the ELP amino acid sequence and the macroscopic properties of the materials. Here we present a review of the structure-property relationships of ELPs and ELP-based materials, with a focus on LCST and mechanical properties and how experimental and computational studies have aided in their understanding.

## 1. Introduction

Materials scientists pay a lot of attention to structural proteins such as elastin, collagen, silk, keratin, or resilin, as a source of nanostructured and multifunctional biopolymers for the production of biomaterials[1]. This is due to their natural abundance, sustainability, multistimuli responsiveness, easy processability, and tuneable structural, mechanical, or optical properties[2,3]. Structural proteins are typically characterised by building blocks formed by short repetitive amino acid motifs[4]. These building blocks can form intricate structures at the nanoscale, such as coiled-coils, β-sheets, or helices, that are key to control the macroscopic mechanical behaviour of the materials derived thereof. Structural proteins can be harvested from natural sources (e.g., animal tissue, silkworm cocoons), but developments in bioprocess engineering, metabolic engineering, and molecular biology are making it increasingly possible to produce them via fermentative processes too[5]. The genetic basis of sequence and length control in proteins provides a unique opportunity to use microbial cultures for the production of rationally designed structural proteins with well-defined chain topologies and interactions. Furthermore, by combining different building blocks, it becomes possible to create *de novo* fusion polypeptides that integrate physicochemical and biological properties of dissimilar proteins into a single biopolymer chain[4].

So far, scaled-up developments in the fermentative production of fusion polypeptides have been hampered by high production costs and low titres, usually on the order of mg/L. These issues are being tackled by optimising bioprocess parameters (e.g., separating growth from the production phase, tuning dissolved oxygen levels or inducer concentration)[6], or genetically engineering the expression microorganism[7]. Furthermore, multiscale computational modelling can support and accelerate the development of structure-property relationships for existing and newly designed structural polypeptides[8]. Materials scientists can use computational simulations to observe *in silico* phenomena with atomic resolution that current analytical techniques cannot reveal and use that data to develop rational guidelines for the selection of polypeptide sequences or processing methods to control the mechanical properties of protein-based materials. Together with developments in manufacturing techniques[9], this is increasing the capacity to manipulate the structure and properties of these materials at the nanoscale.

Elastin is one of the most intriguing structural proteins. A key component of the extracellular matrix (ECM), it provides biological tissues (lungs, blood vessels, ligaments, or skin) with elasticity and resilience. These features, together with the ability to interact with cell receptors and trigger responses like adhesion or chemotaxis, make elastin a very interesting biopolymer e.g., for biomedical applications. Its precursor -the water-soluble tropoelastin- displays a lower critical solution temperature (LCST) behaviour, meaning that it coacervates extracellularly at physiological conditions[10,11]. These coacervates are then cross-linked and integrated into networks of elastic fibres along microfibrillar scaffolding that can withstand indefinite cycles of stress and relaxation. In fact, tropoelastin is able to extend eight times its length and recoil back without hysteresis, making it the most elastic and extensible protein known[12]. Tropoelastin has a molecular weight of 60-72 kDa[13] and is formed by alternating hydrophilic domains rich in lysine and alanine, and repetitive hydrophobic domains rich in valine, proline and glycine, typically with the sequence VPGVG. The low sequence complexity of hydrophobic domains is key to maintain a flexible state in tropoelastin, which together with water interactions assist in conferring elasticity. Conversely, the hydrophilic domains are prone to cross-linking through the lysine residues via lysyl oxidase to enhance its cohesiveness and durability[11,14,15]. Elastin-like polypeptides

(ELPs) consisting of VPGXG repeats display the same LCST behaviour of tropoelastin (although not other properties such as integrin-mediated cell attachment or assembly to make elastin). This means that ELPs are water-soluble below their LCST, but coacervate and phase-separate above it in a reversible process that can take place in temperature intervals as narrow as 1-2 °C[16]. X can be any amino acid except proline[17], and is usually referred to as the guest amino acid. The behaviour of ELPs has inspired many researchers to develop dynamic ELP-based stimuli-responsive materials. Here, we present a review on the progress towards understanding the structure-property relationships of ELPs and ELP-based materials that has been achieved through experimental and computational approaches.

## 2. Experimental studies on the structure-property relationships for ELP-based materials

The difficulty in working with elastin due to its insolubility initially favoured research on ELPs, although there is now a sizeable body of research on the use and application of recombinant tropoelastin too[12,18,19]. Short ELPs were synthesised at first using solid-phase peptide synthesis and studied in solution. However, being able to produce ELPs via bioprocesses[5], using microorganisms like the bacterium *Escherichia coli* or the yeast *Pichia pastoris*, has enabled the development of ELPs with molecular weights much higher than those attainable via solid-phase peptide synthesis. This has made it more accessible to manufacture macroscopic ELP-based materials for applications including biomedicine[20], microfluidics[21], or actuators[22], to name a few. Numerous and varied ELP-based materials have been reported in the form of fibres, films, hydrogels, with a wide range of molecular weights (from 16 kDa[23] to 250 kDa[24]), and physicochemical properties: LCST from <4 °C up to 80 °C[25]; Young's moduli between 30 kPa[26] and 67.4 MPa[24]; or strains at failures between 2%[24] and 1330%[26], as shown in **Figure 1**. Key parameters, such as Young's modulus, ultimate tensile strength, and strain at failure, span up to four orders of magnitude. This showcases that ELPs with disparate properties can be obtained by carefully designing their amino acid sequence length and composition, as well as by controlling their processing conditions into materials.

### 2.1. General properties of ELPs

The coacervation of ELPs upon temperature increases has inspired researchers to uncover the relationships between biopolymer features (i.e., sequence, molecular weight) and the properties of the materials derived from them (i.e., LCST, extensibility, or strength)[27–30]. The seminal work on ELPs dates back to the 1970s[31,32], when short peptides made of combinations of elastin's motif (VPGXG) and mimicking the self-assembling LCST behaviour of elastin were first reported. These polymers were able to coacervate reversibly in response to certain stimuli (e.g., pH, temperature, ionic strength), this responsiveness being determined by the amino acid in the X position. The pioneering work by Urry et al.[32] with the polypentapeptide (VPGXG)$_n$ demonstrated the correlation between the hydrophobicity of the guest amino acid X and the LCST of the peptide[33]. Hydrophobic residues (such as tryptophan, isoleucine, or phenylalanine) reduced the LCST, whereas hydrophilic residues (such as glutamic acid or aspartic acid) increased it. This led the authors to propose a hydrophobicity scale as a function of the guest amino acid X and its mole fraction in the polypeptide. The LCST also decreased by increasing the chain length[34], while charged amino acids in the X position made ELPs responsive to pH[5].

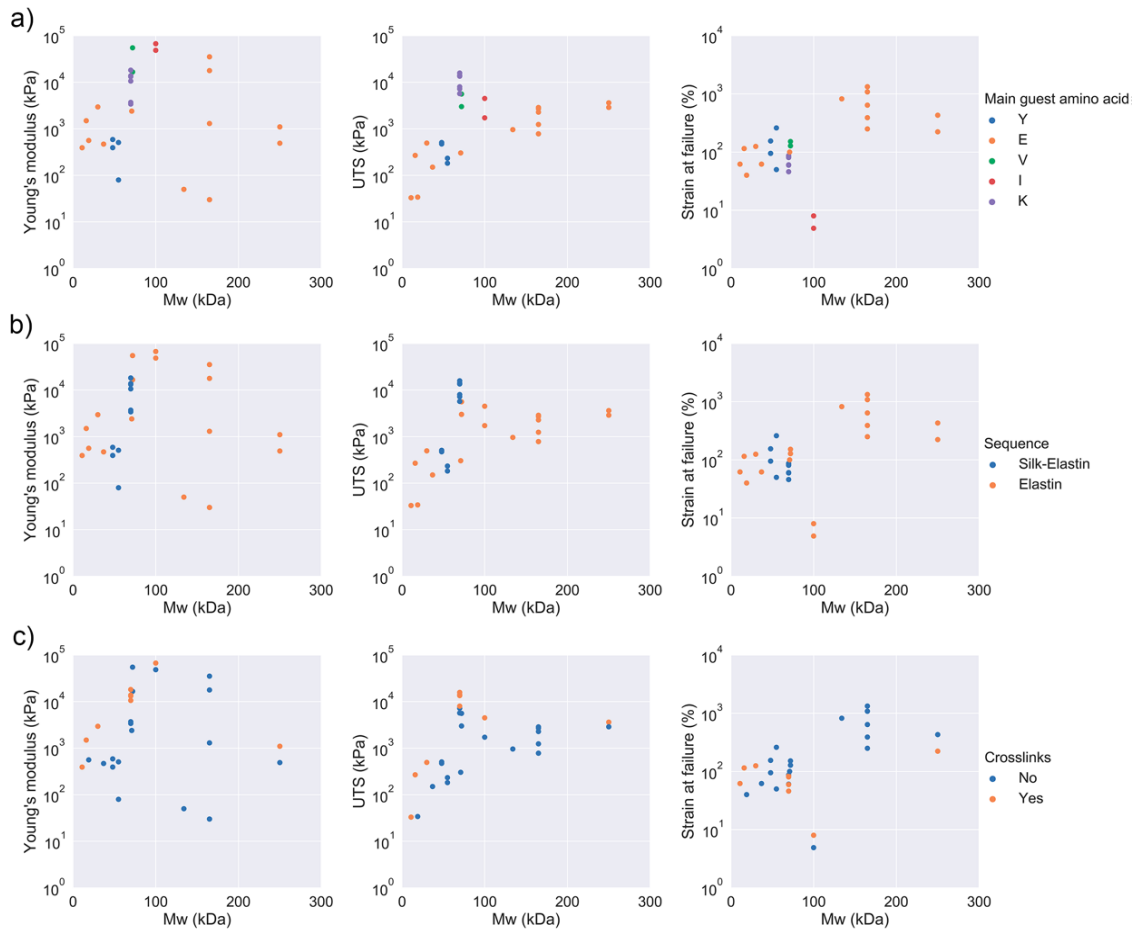

***Figure 1*** – *Young's modulus (in kPa), ultimate tensile strength (UTS, in kPa) and strain at failure (in %) for various ELP-based materials reported in the literature*[24,26,27,73,84,144]*, color-coded by their (a) guest amino acid, (b) presence of cross-links, and (c) functional sequences.*

The LCST and mechanical properties of ELPs and ELPs-based materials can also be modulated using physical and/or chemical cross-links between biopolymer chains[35]. Cross-links can restrict the ability of the chains to undergo structural reshuffling and deswelling above their LCST. Physical cross-links can be achieved using hydrophobic guest amino acids[36], or by fusing ELPs to other physically cross-linking domains like silk-like motifs[37] or coiled coils (discussed in Section 2.3)[38]. However, physical cross-links can be disrupted by external stresses, deeming chemical cross-links a more effective approach to achieve cohesiveness and mechanical strength in ELP-based materials. Chemical cross-links are typically formed by introducing lysine, tyrosine, or cysteine as guest amino acid, and using cross-linking agents like glutaraldehyde, horseradish peroxidase, water peroxide[24,37,39,40], through the reactive polypeptide partners SpyTag and SpyCatcher[41], or via click-chemistry reactions[42–44].

Urry et al. reported that hydrophobic interactions were the driving force for the coacervation of ELPs, and presented a *ΔT mechanism* to describe the molecular basis of the LCST behaviour[33]. They proposed that ordered water molecules surrounding hydrophobic residues in ELPs became less ordered bulk water above the LCST. According to their mechanism, the favourable entropy change of water molecules at higher temperatures exceeded the negative entropy change caused by ELP aggregation, making the coacervation thermodynamically favourable. Other studies reported that increasing concentrations of salt lowered the LCST of ELPs[45] as well. Salt neither interacted through nonspecific electrostatic interactions with ELPs, nor bound to specific groups, because the residues were uncharged and rather nonpolar. Moreover, the effect of salt on ELP assembly and aggregation depends on the type of salt. On the one hand, kosmotropic salts can reduce the LCST by weakening the hydrogen bonds between water and amide groups in the ELP. On the other hand, chaotropic salts can reduce the LCST by destabilising waters hydrating hydrophobic residues. At low concentrations, chaotropic salts can even result in a salting-in effect due to their ability to bind to the ELP backbone[46]. Other studies investigated the changes in the secondary structure above and below the LCST. Initial models proposed that ELPs collapsed above the LCST from a random coil into a more ordered β-spiral structure with three VPGVG units per turn and a β-turn conformation for each pentamer[47–49]. However, our knowledge about the structural implications of the LCST behaviour has evolved since thanks to experimental and computational work[50–53], and the current understanding is that ELPs are disordered below their LCST. Above the LCST, coacervates form by nonspecific, intermolecular hydrophobic contacts, but remain also highly dynamic and disordered.

## 2.2. ELP block copolymers

Understanding the tuneability of the LCST of ELPs by changing e.g., guest amino acid, polymer molecular weight, or polymer concentration, opened the door to design more elaborate biopolymers with complex behaviours. The most common approach has been to develop block copolymers with different guest amino acids that combine dissimilar properties into a single chain (e.g., hydrophilic and hydrophobic blocks; T-responsive and pH-responsive blocks). This has expanded the range of morphologies, stimuli-responsiveness, and mechanical properties of ELP-based materials.

MacKay et al.[54] investigated charged ELPs with basic (X=V/H/G/A [1:2:1:1]) or acidic (X=V/I/E [1:3:1]) guest residues. Charged ELPs were pH-responsive and coacervated only when their charge was neutralised. These and other authors reported quantitative models that predicted the LCST behaviour of charged ELPs as a function of polypeptide sequence, concentration, chain length, or pH[54–56]. ELPs containing pH responsive blocks (with glutamic acid as guest amino acid) showed that higher molecular weights reduced the LCST[57], in a similar fashion to temperature-responsive ELPs. The LCST was determined not only by the mean polarity of the block copolymer but also by the distribution of different polar/nonpolar blocks along the polymer chain[58]: the coacervation temperature increased by reducing the length of the nonpolar block or by fragmenting it and dispersing it between polar blocks throughout the ELP sequence. Incorporating charged amino acids in some of the guest amino acid positions and varying the length of polar blocks in diblock copolymers enabled also the modulation of the high-order supramolecular structures obtained after coacervation (micelles, strings, or physical hydrogels), as well as their mechanical properties[59].

Amphiphilic triblock ELP copolymers with the sequence BAB were also reported[24,60–62]. B represented self-associating hydrophobic end blocks with plastic-like mechanical properties and LCST below physiological temperature, whereas A represented hydrophilic and elastomeric blocks with LCST above 37 °C. Using this design, Nagapudi et al.[26] produced films made of ELP-based thermoplastic elastomers. They designed two amphiphilic triblock copolymers with identical end blocks [(IPAVG)$_4$(VPAVG)]$_n$ separated by a midblock with varying composition. The resulting ELPs had a similar LCST (13-15 °C), despite their wide range of molecular weights (72-165 kDa). They also investigated how casting conditions, including solvent (water or trifluoroethanol), temperature (5-23 °C), and pH (neutral or 14) influenced the mechanical properties of the films. Young's moduli varied between 0.03 and 35 MPa, ultimate tensile strengths between 0.78 and 5.59 MPa, and strains at failure between 250 and 1300%. They suggested that casting aqueous amphiphilic triblock ELP solutions favoured the coacervation of end blocks, yielding a microphase-separated material consisting of islands of hydrophobic end blocks interspersed in solvent-swollen and flexible midblock chains. Stronger materials were obtained using trifluoroethanol (TFE) as a solvent, owing to the better solubility (and thus mixing and entangling) of A and B blocks in TFE that yielded a more interpenetrated polymer network. Other studies with ELP films confirmed the role of the casting solvent on interdomain mixing and its impact on the mechanical properties[63]. Water-cast films subjected to cyclic loading displayed small hysteresis due to the ability of the polymer network to deform and soften stress. Conversely, TFE-cast films displayed viscoplasticity, with large hysteresis, higher viscous losses, and residual strain on cyclic loading, due to the disruption and slippage of biopolymer chains under stress.

Another study investigated five different BAB copolymer constructs to produce films and fibres[27]. Again, end blocks with an identical sequence [(IPAVG)$_4$(VPAVG)]$_n$ flanked central regions with varying sequences. The resulting materials displayed a range of mechanical properties that were comparable to natural hydrogels like hyaluronan or articular cartilage: Young's moduli between 0.03 and 16.65 MPa, tensile strengths between 0.22 and 3.00 MPa, and strains at failure between 128 and 1084 %. Midblocks with the same sequence but different number of repeats showed little change in Young's modulus or ultimate tensile strength, but the strain at failure was significantly higher when using a higher midblock molecular weight. Moreover, these ELPs were successfully subjected to electrospinning. At room temperature, electrospinning was only successful using TFE as solvent. However, by tuning the temperature, aqueous solutions could be electrospun too, enabling the entrapment of bioactive molecules within ELP meshes. A later study described the influence of chemical cross-links on the mechanical properties of films made of ELPs with BAB structure[24]. Two different A midblocks were investigated, containing either isoleucine (hydrophobic) or glutamic acid (hydrophilic) in some of the guest amino acid positions. This had a major influence on the LCST: ELPs with isoleucines in the midblock had a LCST of 21 °C, whereas glutamic acids brought the LCST above 80 °C. Chemical cross-links were introduced using glutaraldehyde to cross-link lysines flanking each of the blocks. These cross-links increased the Young's modulus by a factor of two and reduced the strain at failure by half for both ELPs.

### 2.3. Fusion polypeptides containing ELPs

The functionalities of ELPs can be expanded by fusing them to other peptide sequences (e.g., motifs from silk, collagen, or resilin) into *de novo* recombinant fusion polypeptides, which will be the focus of this subsection. Although outside the scope of this review, ELPs can also be combined with other natural and/or synthetic polymers to create composite materials. Some examples include hyaluronic acide[64], polyethylene glycol[65,66], or collagen microfibres[67].

### 2.3.1. Silk-elastin-like polypeptides

Silk fibroin obtained from the cocoons of the silkworm *Bombyx morii* is one of the toughest materials known, outperforming engineering materials like steel, Kevlar or nylon[1]. This is due to its delicate balance of stiff and strong β-sheet-rich regions (formed by commonly recurring motifs like GAGAGS, GAGAGY, GAGYGA, or GAGAGA), and amorphous regions without distinct repeats that provide elasticity and integrity to silk[68]. The ability to control the formation of β-sheets in silk fibroin materials provides an avenue to control their mechanical properties or their resistance to enzymatic/hydrolytic degradation. Silk-elastin-like polypeptides (SELPs) combine the stiffness and tensile strength of the crystalline domain of silk (usually using the GAGAGS motif) with the elasticity and water solubility of ELPs. SELPs display the LCST behaviour characteristic of ELPs[69], and have been used to synthesise nanoparticles[70], fibres[29,71], free-standing or injectable hydrogels [30,72], or films[29] for applications including tissue scaffolds[73], drug and gene carriers[74–76], or as transfer devices for organoids[77].

Early mechanisms for the self-assembly of SELPs[70] described it as a two-step process: first, micellar-like particles formed, with the silk blocks in the core, driven by hydrophobic interactions between silk domains; second, ELP domains aggregated through hydrophobic interactions. Recent studies proposed an alternative mechanism[78], by which elastin-like domains undergo first an entropy-driven aggregation above the LCST. This aggregation is followed by the self-assembly of silk-like motifs to form β-sheets. According to the latter mechanism, the microstructure of SELP hydrogels is initially dynamic, but becomes arrested over time due to the formation of β-sheets by the silk-like motifs. The ratio between silk-like and elastin-like blocks was key for several properties of SELP materials: hydrogels with a lower silk/elastin ratio were more flexible and had increased responsiveness towards temperature or ionic strength[79]. A higher ratio also decreased the reversibility of the self-assembly of SELPs[70].

The LCST of SELPs can be adjusted in a similar fashion to ELPs: through the selection of the guest amino acid, or by controlling parameters such as biopolymer concentration, molecular weight, or ionic strength[80,81]. A library of 64 SELP sequences with varying composition and molecular weight[69] showed that stimuli including pH, ionic strength, redox, or phosphorylation could also trigger the LCST behaviour. SELPs with valine as guest amino acid had an LCST that decreased with increasing ionic strength or biopolymer concentration in solution. Glutamic acid as guest amino acid rendered the material responsive to pH as well, with LCST increasing as pH rose from 3 to 7. The role of solvent (water, methanol, and formic acid) on the mechanical properties of SELP films was also investigated[29]. After treatment with methanol, SELP films chemically cross-linked using glutaraldehyde exhibited doubled Young's modulus and tensile strength, as well as enhanced extensibility and resilience, compared to non-cross-linked films[82]. Films cast from solutions in water or formic acid were stiff and brittle, but treating them with methanol led to a 10-fold increase in their strain to failure, without significant losses in ultimate tensile strength or Young's modulus. Surfaces in contact with SELP solutions could also impact the morphology of the materials produced. To that end, hydrophilic surfaces with negative potential (e.g., mica) were found to promote the formation of nanofibers[83]. Huang et al. demonstrated the synthesis of free-standing SELP hydrogels with large swelling ratios (up to 84% of the initial weight), reversible changes in optical transparency, and tuneable mechanical properties[37]. SELPs with the same sequence were also used to fabricate thermoresponsive fibres via wet-spinning[84]. Fibres above their LCST could achieve a six-fold increase in their Young's modulus, compared to their values below the LCST, depending on the silk/elastin ratio. SELPs containing valine and lysine as guest amino acids were used to synthesise fibres with high tensile strength and high deformability via electrospinning[85]. SELPs containing arginine, lysine, glutamic acid, and cysteine substitutes were also used to design mucoadhesive drug delivery systems[86]. Cysteine as guest amino acid led to better adhesion, probably due to the formation of disulphide bonds, followed by arginine and lysine, likely because of favourable electrostatic interactions with the negatively charged mucin.

**2.3.2. Collagen-elastin-like polypeptides**

Collagen type I is the most abundant protein in the ECM of mammals[87]. It is characterised by three parallel polypeptide strands that self-assemble into a right-handed triple-helix, forming the tropocollagen molecule[88]. The key motif of triple-helical regions is (G-X-Y)$_n$, proline (P) and 4-hydroxyproline (O) being the most common amino acids in the X and Y positions in mammalian tissues, respectively[89]. 4-hydroxyproline (a post-translational modification of proline) and glycine residues form hydrogen bonds between adjacent chains that stabilise the triple-helix[90]. Tropocollagen can self-assemble into higher-order structures in mammalian tissues, providing mechanical strength, enabling structural organisation of cells and tissue compartments, promoting cell adhesion, and controlling cell differentiation, growth, and pathology[91]. These features have inspired materials scientists to develop collagen-like polypeptides (CLPs) for the fabrication of medical devices[91–93]. CLPs are typically formed by PGP or GPO repeats that mimic the amino acid composition of tropocollagen[43,94,95].

It was proposed that the stability of triple-helical CLPs could modulate the LCST of ELP-CLP materials over a wide range of temperatures. However, most of the combinations of ELPs and CLPs with 4-hydroxyproline residues reported in the literature have been prepared by chemical conjugation, using short CLPs (just 4-6 repeats) synthesised via solid-phase peptide synthesis[25,43,96]. This is due to the challenges associated with obtaining recombinant ELP-CLP polypeptides with post-translational hydroxylation of some proline residues. Even though chemically conjugated ELP-CLP biopolymers are not recombinant fusion polypeptides, we will briefly discuss them, as they provide valuable information for the development of recombinant ELP-CLP biopolymers.

The conjugation of the CLP (GPO)$_4$GFOGER(GPO)$_4$GG with the ELP (VPGFG)$_6$ formed nanovesicles for targeted drug delivery controlled by temperature[43,96]. The melting temperature of CLP was ca. 50 °C, which ensured the formation of a stable triple-helix at physiological temperature. Normally, the short ELP (VPGFG)$_6$ would have a LCST well above physiological temperature, but conjugating it with CLP brought it below 4 °C. This happened in spite of the hydrophilicity of the CLP block, which would *a priori* be deleterious for coacervation. The authors indicated that CLP blocks assembled first to form a triple-helix. ELPs anchored to triple-helical CLPs underwent a local crowding effect, reducing their LCST. These materials showed a high retention on collagen-rich matrices (likely caused by the interaction between collagen triple-helices), as well as cytocompatibility. Recent work demonstrated the high sensitivity of the thermoresponsiveness of ELP-CLP conjugates to the guest amino acids in the ELP block[94]. The CLP (GPO)$_8$GG was conjugated with ELPs with the sequence (VPGXG)$_4$ and containing different ratios of W and F in the X position. W as guest amino acid induced the self-assembly of ELP-CLP conjugates into nanoplatelets at temperatures as low as 4 °C, which was attributed to the higher hydrophobicity of W or the π-π stacking of W side chains. Conversely, the LCST increased beyond 80 °C when F was the main guest amino acid.

### 2.3.3. Resilin-elastin-like polypeptides

Resilin is a highly elastic protein rich in β-turn/β-spiral structures that confers insect wings with their characteristic high resilience and high fatigue lifetime[97]. Two polypeptides inspired in the C- and N-termini of the resilin sequence in *Drosophila melanogaster*[98] are commonly used to create resilin-like polypeptides (RLPs). ELPs and RLPs have in common a high content of P-X$_n$-G motifs, but their different compositions in the X$_n$ positions leads to different mechanical properties: on the one hand, ELPs have a characteristic LCST determined by their content of nonpolar amino acids in the X$_n$ positions; on the other hand, RLPs can display also an upper critical solution temperature

(UCST), meaning that they phase-separate upon cooling below a certain temperature[99]. This UCST behaviour has been attributed to the presence of zwitterionic pairs of residues in the $X_n$ positions, with arginine as the preferred cationic residue[99].

Fusion ELP-RLP biopolymers were synthesised to produce micelles with dual LCST and UCST behaviour[100]. It was shown that the morphology of these micelles could be controlled by tuning the chain length, degree of hydrophilicity, and hydrophilic weight fraction of the ELP block. Moreover, a minimum threshold in the length of the RLP block was identified, below which self-assembly into micelles did not take place. Bracalello et al.[95] developed fusion polypeptides that combined ELPs, RLPs, and CLPs (without hydroxylation of proline residues), and that exhibited a tendency to self-assemble into fibrillar structures. VPGVG ELPs were also fused to the $(GR)_4$ polypeptide, which consisted of a small globular protein GB1 and an RLP[101]. These constructs were used to prepare chemically cross-linked hydrogels that exhibited temperature responsiveness thanks to the ELP block. As a consequence, ELPs coacervated above the LCST, forming additional physical cross-links within the hydrogel structure that led to a higher Young's modulus and a reduced swelling ratio.

### 2.3.4. ELPs fused to other mechanically reinforcing peptides

Several additional peptides can be fused to ELPs to tune their mechanical properties. Some examples include squid ring teeth (SRT) proteins[102] or coiled-coil (C) sequences[38,103,104]. ELPs combined with β-sheet-forming segments from SRT proteins showed the ability to tune by a factor of six the mechanical properties of fibres by controlling the molecular weight of the fusion polypeptide[102]. Coiled-coiled C peptides derived from cartilage oligomeric matrix protein fused to ELPs exhibited very different behaviours above the LCST depending on the position of the C domain. Placing the C domain after the ELP block caused the fusion polypeptides to gel above the LCST, whereas placing the C domain before the ELP block simply turned the solution viscous, without gelling. Leucine zippers (a subtype of coiled-coils) fused to ELPs enabled the formation hydrogels with enhanced durability[38]. While ELPs alone formed micelles above the LCST that dispersed over time, leucine zipper motifs entangled and locked such micelles into more stable structures. ELPs were also fused to leucine zippers and silk-like motifs to develop stable bioinks for 3D printing[103]. By printing above the LCST, the ELP block induced fast gelation based on hydrophobic interactions, followed by stabilisation by the leucine zippers. This fast aggregation enabled a progressive formation of β-sheets between silk-like motifs to consolidate the 3D printed structures.

### 2.3.5. ELPs fused to biomineralising peptides

Living organisms can express peptides that nucleate the growth of mineral structures, as seen in materials like bone, nacre, or teeth[105]. These peptides can be fused to ELPs to develop mineralised materials with applications e.g., bone regeneration. One example of such peptides is derived from the N-terminus of the salivary protein statherin ($SN_A15$), which can promote the growth of calcium phosphate structures[106]. To that end, $SN_A15$ was fused to ELPs containing isoleucine and lysine as guest amino acids. ELP-$SN_A15$ membranes showed an enhanced mineralisation of calcium phosphates, as well as *in vivo* bone regeneration properties[107,108], but successful mineralisation was only achieved above a minimum content of $SN_A15$ blocks[109]. Moreover, the size and shape of calcium phosphate structures could be tuned by modifying the balance between ordered and disordered regions in ELPs (controlled by drying the solution or by forming chemical cross-links)[110]. The size and shape of the minerals also affected the mechanical properties of the resulting materials[108,110], which in some cases attained a Young's modulus larger than those from bone or dentin. ELPs were also mineralised with silica by fusing them to the N-terminal silaffin R5 peptide of the diatom species *Cylindrotheca fusiformis*[111]. In a first step, the fusion polypeptide formed micellar assemblies via the ELP domains that were conjugated to several hydrophobic drugs via a cysteine-rich C-terminal trailer. Subsequently, the R5 peptide drove the polycondensation of the silica-precursor tetramethylorthosilicate to form uniform, hybrid core–shell silica nanoparticles encapsulating the ELP-drug conjugates.

**2.3.6. ELPs fused to bioactive domains**

ELPs can also be fused to domains that modulate the biological properties of ELP-based materials. Some examples include cell-binding sequences such as RGD[66,108] (for unspecific cell adhesion) or REDV[107,112,113] (for endothelial cell adhesion); the incorporation of protease sensitive sequences[113,114]; or fusing ELPs with antimicrobial peptides[115]. ELPs containing the sequence REDV and chemically cross-linked via lysines exhibited increased adhesion and spreading of endothelial cells when the lysine residues were confined to the terminal regions of the biopolymer chains[113]. This was attributed to conformational changes in the biopolymer that influenced the accessibility of the REDV sequence or its affinity for the $\alpha 4\beta 1$ integrin receptor on the endothelial cell surface. This work showed that amino acids located 15 or more residues away from the REDV sequence could still impact the adhesion and spreading of endothelial cells. Being able to program the degradation of ELP-based materials through its amino acid sequence is a particularly interesting aspect for biomedical applications. To that end, the insertion of the VGVAPG protease recognition site for human elastase enabled complete degradation of ELPs by enzymatic digestion, generating fragments with sequences similar to those obtained from natural elastin during the ECM rearrangement[113]. SELPs were also fused to a matrix-metalloproteinase (MMP) sensitive peptide. This enabled the synthesis of hydrogels with controlled degradation for the localised delivery of bioactive agents in tumour areas, since several types of solid tumours have high levels of MMPs[114].

**3. Multiscale computational modelling studies on the structure-property relationships for tropoelastin and elastin-like polypeptides**

The number of possible ELP sequences makes their design space so vast as to become almost intractable via just experimental work. To that end, multiscale computational modelling can be used to investigate *in silico* the mechanical properties of ELPs and ELP-based materials across different length scales, from the amino acid sequence to the macroscale, reducing the number and cost of time-consuming experimental biosynthetic trials[116]. The following section highlights how different multiscale computational modelling techniques have contributed to the understanding and the prediction of the structure-property relationships for tropoelastin, ELPs, and ELP-based materials.

Computational models with various levels of resolution can be developed, depending on the length- and time-scales at which the system is modelled, spanning quantum mechanical methods, full atomistic and coarse grained (CG) molecular dynamics (MD) methods, up to finite elements methods. Parameters like solvent, ionic strength, or processing options applied can be included in the models. Computational models have been applied mostly to complement experimental work, with only a limited set of studies taking advantage of their predictive power to design new ELP-based materials[117,118]. This is caused by challenges in the implementation of computational models, especially related to insufficient computational power to access time- and length-scales that reproduce some of the key phenomena for ELPs (e.g., coacervation). Nonetheless, the growth in computational capacity and the development of advanced sampling techniques and machine learning methods will continue to expand the computational tools available to accelerate the development of ELP-based materials[119].

Full atomistic MD simulations provide information about protein dynamics. These simulations calculate numerical solutions for classical many-body problems in which atoms are considered as point masses that interact with each other following Newtonian equations of motion. Full atomistic MD simulations have been successfully applied to many biomolecular systems[37,120–122]. To that end, force fields that accurately calculate the forces and energies in these systems are needed. Some examples of extensively validated force fields for biomolecular simulations include CHARMM[117,123,124], AMBER[125–127], or OPLS[128–131]. The secondary structure of proteins is key to the mechanical properties, but predicting it for *de novo* polypeptide designs via full atomistic MD simulations is challenging, especially if solvent molecules are explicitly represented in the simulation. Generally, the time scales of their folding (starting from an idealised straight chain) are beyond reach for full atomistic MD simulations, which can simulate only a few hundred nanoseconds. In those cases, one has to either resort to advanced sampling methods, or sacrifice atomistic resolution and apply CG simulations. Replica-based sampling methods, such as replica exchange molecular dynamics (REMD)[132] can accelerate the prediction of the protein secondary structure by focusing on conformational sampling, albeit destroying dynamical information. Another option is to resort to CG models, which are computationally more economical (at the expense of atomistic resolution), and allow exploration of length- and time-scales inaccessible via full atomistic MD simulations. Common CG methodologies include dissipative particle dynamics (DPD)[133] or the MARTINI force field[134].

## 3.1. Tropoelastin

The structure of human elastin remained elusive for decades, due to the difficulty in isolating tropoelastin monomers, although this has since been resolved by using recombinant expression of human tropoelastin[12,18,19]. Tropoelastin was initially considered an intrinsically disordered protein, but recent studies proved that it has a defined nanostructure in solution[12] that balances disordered regions (needed for molecular elasticity) with regions with a defined structure that determine its self-assembly into elastic fibres[124]. Multiscale computational modelling facilitated the uncovering of tropoelastin's nanostructure, as well as the molecular basis of some diseases caused by mutations in its sequence[135]. Yeo et al. used REMD to study how changes at the molecular scale controlled the order-disorder balance in tropoelastin[124], demonstrating that minor changes in tropoelastin (<4% of the entire sequence) prevented the formation of elastin fibres. This was caused by both local changes (a large fraction of flexible helices was replaced by more rigid β-sheets) and global changes in the molecular shape. Tarakanova et al.[135] developed a full atomistic model for human tropoelastin using REMD. Their structure, validated against experimental data, showed the amenability of REMD simulations to predict the structure of proteins with highly disordered regions. The tropoelastin monomer consisted of three main regions: an elastic region close to the N-terminus, a foot-like region containing cell-interacting domains, and a connecting bridge that modulated fibre self-assembly. The canonical shape of tropoelastin was generally maintained at 37 °C, despite the flexibility of some of its regions[136]. Specifically, lysines displayed an enhanced mobility compared to other amino acids. Lysines are involved in the formation of intermolecular cross-links between tropoelastin monomers, which is a key step towards the formation of elastin fibres[11]. The formation of such cross-links is preceded by the modification of lysine into allysine, which was shown in simulations to change the canonical shape of tropoelastin and facilitate the formation of intermolecular cross-links[137].

The aforementioned studies used full atomistic MD simulations. However, the coacervation of several tropoelastin monomers and the formation of elastin fibres is a process beyond the time-scales accessible by full atomistic MD simulations. Thus, a CG model was developed using the MARTINI force field[138]. Ten microsecond simulation revealed that tropoelastin coacervation is a temperature-dependent stepwise process, initiated by a nucleation step with intermolecular interactions through hydrophobic domains and followed by a fibre growth step. This study highlighted the potential for CG MD simulations to study the temperature-dependence of the coacervation of several tropoelastin monomers to form multimeric higher-order structures.

### 3.2. Elastin-like polypeptides

Full atomistic MD simulations (mainly using variations of the VPGVG motif with a reduced number of repeats) were used to investigate the secondary structure of ELPs above and below the LCST. Although initial models proposed an idealised β-spiral structure for ELPs above the LCST[47–49], MD simulations and experimental data demonstrated that ELPs are rather flexible molecules both above and below the LCST, with local, short-termed conformational preferences towards β-turns[50,51,53,139]. Furthermore, MD simulations revealed a threshold in proline and glycine content key to maintaining the ELP backbone hydrated and disordered even after coacervation. This prevented the hydrophobic collapse of the backbone because of its inability to form extensive hydrogen-bonded, water-excluding self-interactions[128]. Overall, computational simulations have shown that a certain level of disorder is an essential requirement for the self-assembly of ELPs.

Most of the MD work investigated the molecular basis of the LCST behaviour using computational systems with one polypeptide chain. The role of water in the coacervation was studied, albeit with a short simulation time (6-9 ns)[51]. The ability of water to form hydrogen bonds with the backbone indicated that ELPs were fully dynamic systems. Pulling/releasing MD simulations with the ELP (VPGVG)$_{18}$ showed that the entropy of water molecules hydrating hydrophobic amino acids was also responsible for the elastic recoil of ELPs[50]. Despite the dynamic and hydrated character of ELPs, a decrease in the peptide/water interactions was reported above the LCST, which was attributed to a hydrophobic collapse[51]. Longer MD simulations with 32 ns with the ELP GVG(VPGVG) arrived at similar conclusions[123,140]: the authors indicated that short chains behaved like a two-state system of extended and folded structures in a dynamic equilibrium. The LCST marked a sharp variation in the relative population of each of the states, with a decrease in the peptide-water interactions and a speedup of the peptide backbone dynamics above the LCST. A follow-up study testing three different force fields (CHARMM, GROMOS and OPLS), two solvent modes (TIPS3P and SPC), and longer simulation times (between 60 and 140 ns) showed qualitative agreement with previous studies[141].

Conversely, extended MD simulations (350 ns) with the ELP GVG(VPGVG)$_3$[126] showed a different result and did not identify any LCST behaviour. This led the authors to challenge the idea of the folding of ELPs above their LCST. In fact, ELPs at low temperature appeared to be more rigid due to stabilisation by water molecules, whereas higher temperatures promoted the thermal breaking of the network of hydrating water, favouring a higher disorder. Zhao et al.[142] also did not find evidence of a structural transition above the LCST for short chains with just 3 repeats of the VPGVG motif but did observe it for chains with 10, 18, and 30 repeats. This hinted at the inability of classical MD simulations to reproduce the LCST behaviour for very short ELPs. To circumvent this issue, Tarakanova et al.[117] applied REMD to investigate the effect of chain length, salt addition and guest amino acid on the LCST of ELPs with just 3-6 repeats. The results showed a structural collapse with just three repeats, although longer ELPs (with six repeats) favoured the expulsion of water molecules from the polypeptide bulk to form a more compact structure with a higher content of β-turns. Higher salt concentrations lowered the LCST because of the stronger interaction between salt ions and electronegative ions in the peptide chain, disrupting hydrogen-bonded water networks around the peptide. Moreover, valine as guest amino acid enhanced the collapse above the LCST, due to its higher hydrophobicity, whereas lysine stiffened the ELP hinge due to its stronger interaction with water, hindering the aggregation. These results were validated experimentally, showing the predictive power of REMD to predict the LCST of short ELPs.

The work described so far reported simulations of one single ELP chain. However, multichain simulations are necessary to isolate single-chain effects from collective events during coacervation. To that end, Li et al.[127] reported MD simulations (>70 ns) using two $(VPGVG)_{18}$ chains. They reported that higher temperatures exposed hydrophobic valine side chains to the solvent, which drove coacervation and formed more β-turns. The impact of sequence directionality on the LCST behaviour was also investigated for $(VPGVG)_{18}$ and $(VGPVG)_{18}$[23] in one- and two-chain systems. The results, validated against experimental data, showed that the LCST behaviour of $(VPGVG)_{18}$ was reversible, whereas $(VGPVG)_{18}$ coacervates remained aggregated even after significant undercooling. This was attributed to the higher surface hydrophilicity observed for $(VGPVG)_{18}$ above its LCST, which caused stronger interchain interactions.

The seemingly contradictory results on the presence or absence of order in ELPs above their LCST was resolved by the most extensive MD simulation reported to date (combined sampling time exceeding 200 μs), using systems consisting of 1 and 27 ELP chains[52]. ELPs displayed disordered structural ensembles both as individual chains or in aggregate, with no preferred conformation, although individual chains showed a larger collapse above the LCST than the multichain system. ELPs remained hydrated even after coacervation, with each chain approaching their maximally-disordered, melt-like structure ("liquid protein state"). The 27-chain system was stabilised by non-specific hydrophobic interchain contacts above the LCST, but without the formation of a hydrophobic core. The secondary structure (mainly in the form of β-turns) was sparse and fluctuating, with a duration in the sub-nanosecond scale, and without major changes upon coacervation. This work reconciled observations of highly disordered ELP chains before and after coacervation, and local secondary structure in the form of β-turns.

### 3.3. Fusion polypeptides containing ELPs

Several computational studies investigated fusion and chemically-conjugated polypeptides that merged ELPs with silk-like polypeptides[37,84,118,143], collagen-like polypeptides[25,131], or the LG5 domain of the α2 chain from laminin[120]. However, the body of computational work with fusion polypeptides is still scarce due to the aforementioned challenges pertaining simulations of long polypeptides. MD simulations with SELPs studied their unfolding process under tensile stress[143], using different pulling speeds above and below their LCST. Similar to studies with ELPs at the single molecule level[52,117], they reported a collapse of SELPs above the LCST using REMD. The collapse was accompanied by an increase in the number of intramolecular hydrogen bonds. Pulling of these polypeptides using steered MD simulations[84,143] revealed the temperature-dependent nanomechanics of SELPs. Higher energy dissipation was observed when pulling above the LCST. This was counterintuitive, as easier unravelling would normally be expected from materials at a higher temperature. CG models were also developed for SELPs using the PLUM potential[23]. These CG models (validated against experimental data and full atomistic MD) were able to reproduce the LCST behaviour, as well as the influence of biopolymer concentration or chemical cross-linking on the LCST of SELPs hydrogels[118]. This showcased the predictive power of CG models to design, optimise, and customise SELP biomaterials.

Simulations on CLP-ELP polypeptides[131] were performed to study the experimentally observed shift to lower LCSTs for certain ELPs when conjugated with CLPs[96]. No chain collapse or phase separation was observed in full atomistic MD simulations, likely due to the short ELP length. Thus, a CG model was developed using a generic bead-spring polymer model to explore extended time- and length-scales. The CG model reproduced the lower LCST of ELPs when anchored to CLPs. Subsequently, Prhashanna et al.[25] investigated ELP-CLP conjugates with phenylalanine or tryptophan as guest amino acids. Their simulations showed that ELPs with W substitutes were more prone to acquire β-turn structures at lower temperatures than with F substitutes.

Laminins are a family of ubiquitous extracellular proteins, found in the basal lamina of almost every tissue. This makes them an attractive bioactive component to fuse with ELPs, with the goal of promoting assembly and structural support for the development of tissue engineering scaffolds or drug delivery devices. Full atomistic MD simulations showed that ELPs fused to the LG5 domain of the α2 chain from laminin[120] behaved similarly to free ELPs, without interacting with the laminin domain, and with similar water dynamics and propensity towards the formation of β-turn structures. The authors claimed that their methodology could be applied to simulate multiple fusion protein designs to rank-order them before synthesising the best-performing candidates for the production of engineered ECM proteins, although no experimental validation of the computational work was provided in that study.

## 4. Outlook

Elastin has been a source of inspiration for the development of nanostructured and dynamically responsive multifunctional materials since the early work on short and thermoresponsive elastin-like polypeptides in the 1970s. This has triggered research to create lightweight and elastic yet mechanically robust ELP-based materials with responsiveness towards environmental variables (i.e., temperature, salt concentration, or pH). The properties of these materials can be expanded by fusing them to other peptides, such as collagen-like polypeptides, silk-like motifs, biomineralising sequences, or resilin-like polypeptides. This has become increasingly possible thanks to advances in molecular biology, bioprocess engineering, and new material manufacturing techniques, bringing along opportunities for the synthesis of new ELP-based materials with controlled nanostructure, stimuli-responsiveness, or mechanical properties. The rational design of such new materials can benefit from the use of computational modelling tools that aid in the development of structure-property relationships for ELPs. Such tools have proven useful in elucidating fundamental properties of ELPs (e.g., the molecular basis of coacervation or the effect of LCST on their secondary structure), as well as in providing guidelines for the synthesis of ELP-based materials. Overall, structural biopolymers in general and ELPs in particular are an emerging and promising field for the development smart materials for various applications, including minimally invasive drug delivery, tissue regeneration, or soft robotics.

**Acknowledgements**

This project has received funding from the European Union's Horizon 2020 research and innovation program under the Marie Skłodowska-Curie Grant Agreement SUPERB 892369.